\documentclass[12pt]{iopart}
\usepackage{amssymb}
\usepackage{graphicx}
\usepackage{amsthm}
\usepackage{psfrag}
\usepackage{dsfont}

\newtheoremstyle{custom}
{15pt} {15pt} {} {} {\bfseries} {} {.5em} {} \theoremstyle{custom}
\newtheorem{definition}{Definition}

\newtheorem{lemma}[definition]{Lemma}
\newtheorem{corollary}[definition]{Corollary}
\newtheorem{theorem}[definition]{Theorem}

\newtheorem{condition}[definition]{Condition}
  
\begin{document}

\title{Conformal Structures Admitted by a Class of FRW Cosmologies}
\author{Philip Threlfall and Susan M Scott}
\address{Centre for Gravitational Physics, \\
College of Physical Sciences, \\
The Australian National University,\\
Canberra ACT 0200\\
Australia} \eads{\mailto{phil.threlfall@anu.edu.au},
\mailto{susan.scott@anu.edu.au}}

\begin{abstract}
In this paper we demonstrate that there are large classes of
Friedmann-Robertson-Walker (FRW) cosmologies that admit isotropic
conformal structures of Quiescent Cosmology. FRW models have long
been known to admit singularities such as Big Bangs and Big Crunches
\cite{Lazkoz, Visser} but recently it has been shown that there are
other cosmological structures that these solutions contain. These
structures are Big Rips, Sudden Singularities and Extremality Events
\cite{Lazkoz, Visser}. Within the Quiescent Cosmology framework
\cite{Barrow} there also exist structures consistent with a
cosmological singularity known as the Isotropic Past Singularity
(IPS) \cite{Scott, Goode}. There also exists a cosmological final
state known as a Future Isotropic Universe (FIU) \cite{Scott}, which
strictly speaking, doesn't fit with the fundamental ideals of
Quiescent Cosmology. In this paper, we compare the cosmological
events of a large class of FRW solutions to the conformal structures
of Quiescent Cosmology \cite{Scott}.\\

In the first section of this paper we present the relevant
background information and our motivation. In the second section of
this paper we construct conformal relationships for relevant FRW
models. The third section contains a thorough discussion of a class
of FRW solutions that cannot represent any of the previously
constructed isotropic conformal structures from Quiescent Cosmology.
The final section contains our remarks and future outlook for
further study of this field.
\end{abstract}
\maketitle
\section{Background and Motivation}
\subsection{The FRW Cosmologies}
The FRW cosmologies are the simplest, non-trivial solutions to the
Einstein field equations that represent homogeneous and isotropic
cosmologies which contract or expand \cite{Ellis}. They are of
fundamental importance because, although simple, they (at least
approximately) describe the physical properties of the observable
Universe. The FRW metric is given, in spherical polar coordinates
$(t, r, \theta, \phi)$, by
\begin{eqnarray}
\rmd s^{2} &=& -\rmd t^{2} + a^{2}(t)\left[\frac{\rmd
r^{2}}{1-kr^{2}} + r^{2}\left(\rmd\theta^{2} + \sin^{2}\theta
\rmd\phi^{2}\right)\right]\textrm{,}\label{FRWMetric}
\end{eqnarray}
where $a(t)$ is known as the scale factor and $k = -1$, $0$, $+1$
represents the curvature of the 3-space.\\

The most well-known types of cosmological events present in the FRW
space-times are the singularities known as the Big Bang if the
singularity is in the past and the Big Crunch if the singularity is
in the future. In both cases the scale factor goes to zero. Recently
there has been considerable interest in a new class of cosmological
events, including Big Rips, Sudden Singularities and Extremality
Events \cite{Lazkoz, Visser, Caldwell1, Caldwell2, Barrow2}.
\subsection{Cosmological Events in the FRW
Framework}\label{CosmologicalEvents} When considering cosmological
events it is instructive to consider the scale factor as a
generalised Puiseux series expansion around the time, $t_{0}$, when
the event occurs, as seen in \cite{Lazkoz} and \cite{Visser},
\begin{eqnarray}
a(t) &=& c_{0}|t-t_{0}|^{\eta_{0}} + c_{1}|t-t_{0}|^{\eta_{1}} +
c_{2}|t-t_{0}|^{\eta_{2}} + \cdots\textrm{,}
\end{eqnarray}
where the exponents $\eta_{i}\in\mathbb{R}$ are ordered such that
$\eta_{i} < \eta_{i+1}$. There are no restrictions on the constants,
$c_{i}$, apart from $c_{0} > 0$ and $c_{i} \neq 0$. In the following
subsection we will adopt the definitions used in the comprehensive
paper by Catto\"en and Visser \cite{Visser} but, in the interests of
simplicity, we will suppose that all cosmological events happen at a
time $t_{0}$.
\subsubsection{Big Bangs/Big Crunches}
A cosmological event is said to be a Big Bang (Big Crunch) if $a(t)
\rightarrow 0$ at some finite time $t_{0}$ in the past (future). The
scale factors for these singularities will be expressed as
\begin{eqnarray}
a(t) &=& c_{0}(t-t_{0})^{\eta_{0}} + c_{1}(t-t_{0})^{\eta_{1}} +
c_{2}(t-t_{0})^{\eta_{2}} + \cdots\textrm{,}
\end{eqnarray}
for a Big Bang and
\begin{eqnarray}
a(t) &=& c_{0}(t_{0}-t)^{\eta_{0}} + c_{1}(t_{0}-t)^{\eta_{1}} +
c_{2}(t_{0}-t)^{\eta_{2}} + \cdots\textrm{,}
\end{eqnarray}
for a Big Crunch. The exponents will satisfy $0 < \eta_{0} <
\eta_{1} < \eta_{2} < \cdots$ and $\eta_{i} \in \mathbb{R}^{+}$.
\subsubsection{Big Rips}
In the opposite cosmological scenario, we can have the scale factor
becoming infinite as $t_{0}$ is approached to the future. In this
case the cosmological event is referred to as a Big Rip. The scale
factor has a form similar to that of the Big Crunch but in this case
$\eta_{0} < 0$ and $\eta_{0} < \eta_{1} < \cdots$ :
\begin{eqnarray}
a(t) &=& c_{0}(t_{0}-t)^{\eta_{0}} + c_{1}(t_{0}-t)^{\eta_{1}} +
c_{2}(t_{0}-t)^{\eta_{2}} + \cdots\textrm{,}
\end{eqnarray}
with $c_{0} > 0$ and $\eta_{i} \in \mathbb{R}$.
\subsubsection{Sudden Singularities}
We have considered zero and infinite scale factors but we now turn
our attention to a finite non-zero scale factor at $t_{0}$. If the
scale factor is finite but one of its time derivatives diverges then
we will say that the cosmological event is a Sudden Singularity. The
scale factor can be written as
\begin{eqnarray}
a(t) &=& c_{0} + c_{1}|t_{0}-t|^{\eta_{1}} +
c_{2}|t_{0}-t|^{\eta_{2}} + \cdots\textrm{,}
\end{eqnarray}
where we have set $\eta_{0} = 0$ and $0 < \eta_{1}< \eta_{2}\cdots$,
$\eta_{i}\in\mathbb{R}^{+}$. We also require, for some $i$, with
$\eta_{i} \notin \mathbb{N}$, that a time derivative,
\begin{eqnarray}
a^{n}(t)&\sim& c_{i}\eta_{i}(\eta_{i} - 1)\cdots(\eta_{i} - n +
1)|t_{0} - t|^{\eta_{i} - n}\textrm{,}
\end{eqnarray}
diverges as $t\rightarrow t_{0}$.
\subsubsection{Extremality Events}
This class of events does not contain any singularities. For
Extremality Events the scale factor $a(t)$ exhibits a local extremum
at finite time $t_{0}$; in particular $a'(t)\rightarrow 0$ as
$t\rightarrow t_{0}$. It has the same form as the scale factor for
Sudden Singularities but $\eta_{i} \in \mathbb{N}$ for all $i>0$.
\begin{itemize}
\item A `bounce' is any local minimum of $a(t)$ such that $a'(t_{0}) =
0$. The `order' of the bounce is the first non-zero integer $n$ for
which
\begin{eqnarray}
a^{(2n)}(t_{0}) > 0\textrm{,}
\end{eqnarray}
so that
\begin{eqnarray} a(t) &=& a(t_{0}) +
\frac{1}{(2n)!}a^{(2n)}(t_{0})[t_{0}-t]^{2n} + \cdots\textrm{ .}
\end{eqnarray}
\item A `turnaround' is any local maximum of $a(t)$ such that $a'(t_{0}) =
0$. The `order' of the turnaround is the first non-zero integer $n$
for which
\begin{eqnarray}
a^{(2n)}(t_{0}) < 0\textrm{,}
\end{eqnarray}
so that
\begin{eqnarray}
a(t) &=& a(t_{0}) + \frac{1}{(2n)!}a^{(2n)}(t_{0})[t_{0}-t]^{2n} +
\cdots\textrm{ .}
\end{eqnarray}
\item An `inflexion event' is an Extremality Event that is
    neither a local minimum nor a local maximum. The `order' of
    an inflexion event is the first non-zero integer $n$ for
    which
\begin{eqnarray}
a^{(2n + 1)}(t_{0}) \neq 0\textrm{,}
\end{eqnarray}
so that
\begin{eqnarray}a(t) &=& a(t_{0}) + \frac{1}{(2n + 1)!}a^{(2n
+ 1)}(t_{0})[t_{0}-t]^{2n + 1} + \cdots\textrm{ .}
\end{eqnarray}
\end{itemize}
\subsection{Quiescent Cosmology}
According to classical General Relativity, it is generally agreed
that the Universe we observe today started with a Big Bang, there
remains some debate as to the order of the primordial quantum foam
from which the Universe was forged. It can be argued that the
Universe was highly ordered in the beginning or that it was highly
disordered. The former leads to the formulation of Quiescent
Cosmology \cite{Barrow}, the latter to Chaotic Cosmology
\cite{Misner}. We shall concern ourselves here with Quiescent
Cosmology and consider the associated conformal
relationships of the metric.\\

Quiescent Cosmology was introduced by Barrow in 1978 \cite{Barrow}
and effectively states that the Universe began in a highly ordered
state and has evolved away from its highly regular and smooth
beginning because of gravitational attraction. We continue to see
isotropy on large scales because we exist in an early stage of
cosmological evolution. The kind of initial cosmological singularity
that is compatible with the Quiescent Cosmology concept is one that
is isotropic. This type of initial isotropic singularity was given a
rigorous mathematical definition in 1985 by Goode and Wainwright
\cite{Goode} using a conformal relationship of the metric. The
version of the definition given below is due to Scott \cite{Scott2}
who removed the inherent technical
redundancies of the original definition.\\

Goode and Wainwright were concerned with the past of the Universe
(Big Bang). In recent years, however, there has been increasing
interest in possible future evolutions of the Universe and its
eventual fate. In 2009 H\"ohn and Scott \cite{Scott} investigated
possible isotropic and anisotropic future end states of the Universe
and introduced new definitions to cover the various possible
scenarios. Following Goode and Wainwright they also employed
conformal relationships of the metric.
\subsection{Conformal Structures}
\begin{figure}[h!]
\begin{center}
\psfrag{A}{ } \psfrag{B}{ }
\psfrag{C}{$\begin{array}{l} \mathbf{g} = \Omega^{2}\left(T\right)\tilde{\mathbf{g}}\\
\Omega\left(0\right) = 0\\
\end{array}$}
\psfrag{D}{$\mathbf{u}$} \psfrag{E}{$\tilde{\mathbf{u}}$}
\psfrag{F}{$T = 0$} \psfrag{G}{$T = 0$} \psfrag{H}{$\begin{array}{l}\textrm{Physical space-time}\\
\left(\mathcal{M}, \mathbf{g}\right)\end{array}$}
\psfrag{I}{$\begin{array}{l}\textrm{Unphysical space-time}\\
\left(\tilde{\mathcal{M}}, \tilde{\mathbf{g}}\right)\end{array}$}
\caption{A pictorial interpretation of an IPS. The fluid flow is
represented by $\mathbf{u}$.\\
\\
}\label{IPS}
\includegraphics[scale = 0.35]{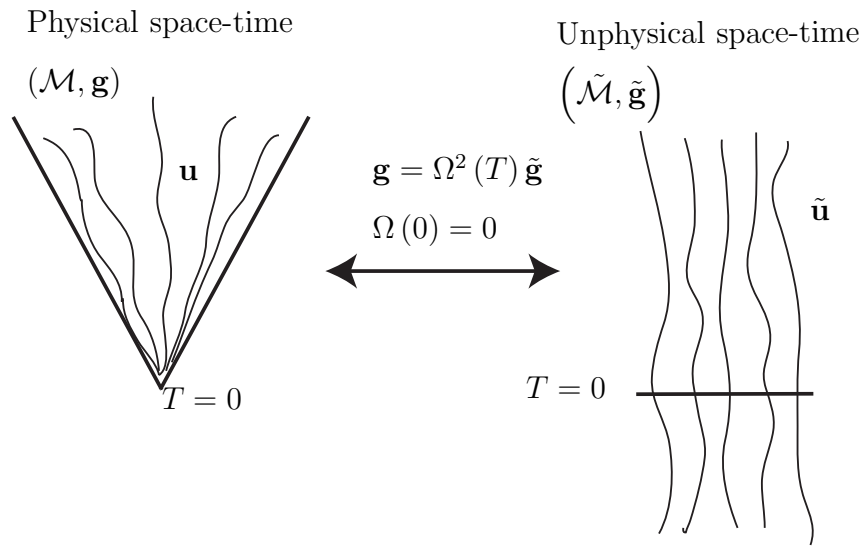}
\end{center}
\end{figure}

\begin{figure}[h!]
\begin{center}
 \psfrag{A}{$\bar{T} = 0$} \psfrag{B}{$\bar{T} = 0$}
\psfrag{C}{$\begin{array}{l} \mathbf{g} = \bar{\Omega}^{2}\left(\bar{T}\right)\bar{\mathbf{g}}\\
\bar{\Omega}\left(0\right) = 0\\
\end{array}$}
\psfrag{D}{ } \psfrag{E}{ }
\psfrag{F}{$\begin{array}{l}\textrm{Physical space-time}\\
\left(\mathcal{M}, \mathbf{g}\right)\end{array}$} \psfrag{G}{$\begin{array}{l}\textrm{Unphysical space-time}\\
\left(\bar{\mathcal{M}}, \bar{\mathbf{g}}\right)\end{array}$}
\psfrag{H}{$\mathbf{u}$} \psfrag{I}{$\bar{\mathbf{u}}$} \caption{A
pictorial interpretation of an IFS. The fluid flow is represented by
$\mathbf{u}$. It can be seen that an IFS is essentially a time
reversal of an IPS.\\
\\}
\includegraphics[scale = 0.35]{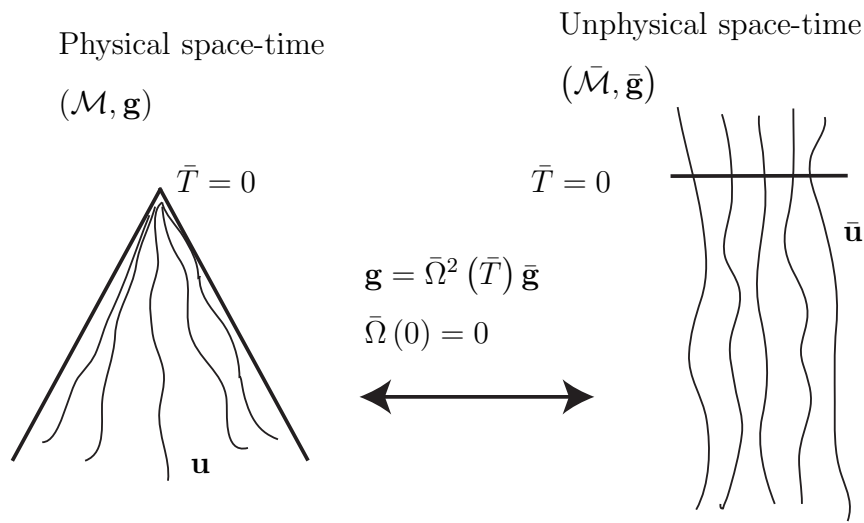}
\label{IFS}
\end{center}
\end{figure}

Since, in this paper, we are dealing with FRW space-times which are
homogeneous and isotropic, we will present here only the conformal
definitions that relate to isotropic initial and final states of the
Universe. These will comprise the Isotropic Past Singularity, the
Isotropic Future Singularity and the Future Isotropic Universe. Any
ancillary definitions that are needed will also be included.
\begin{definition}[Conformally related metric]
A metric $\mathbf{g}$ is said to be conformally related to a metric
$\mathbf{\tilde{g}}$ on a manifold $\mathcal{M}$ if there exists a
conformal factor $\Omega$ such that
\begin{eqnarray}
\mathbf{g} &=& \Omega^{2}\mathbf{\tilde{g}}\textrm{, where $\Omega$
is a strictly positive function on $\mathcal{M}$.}
\end{eqnarray}
\end{definition}
\begin{definition}[Cosmic time function]
For a space-time $(\mathcal{M},g)$, a cosmic time function is a
function $T$ on the manifold $\mathcal{M}$ which increases along
every future-directed causal curve.
\end{definition}
It should be noted that we will henceforth denote relevant
quantities for past cosmological frameworks with a tilde $(\sim)$
and for future cosmological frameworks we will use a bar $(-)$.
\begin{definition}[Isotropic Past Singularity (IPS)]
A space-time $(\mathcal{M},\mathbf{g})$ admits an Isotropic Past
Singularity if there exists a space-time $(\tilde{\mathcal{M}},
\mathbf{\tilde{g}})$, a smooth cosmic time function $T$ defined on
$\tilde{\mathcal{M}}$ and a conformal factor $\Omega(T)$ which
satisfy
\begin{enumerate}
\item[i)] $\mathcal{M}$ is the open submanifold $T > 0$,
\item[ii)] $\mathbf{g} = \Omega^{2}(T)\mathbf{\tilde{g}}$ on
$\mathcal{M}$, with $\mathbf{\tilde{g}}$ regular (at least $C^{3}$
and non-degenerate) on an open neighbourhood of $T = 0$,
\item[iii)] $\Omega(0) = 0$ and $\exists\; b > 0$ such that $\Omega \in C^{0}[0,b]
\cap C^{3}(0,b]$ and $\Omega(0,b] > 0$,
\item[iv)] $\lambda \equiv \lim\limits_{T\rightarrow 0^{+}}L(T)$ exists, $\lambda \neq
1$, where $L \equiv
\frac{\Omega''}{\Omega}\left(\frac{\Omega}{\Omega'}\right)^{2}$ and
a prime denotes differentiation with respect to $T$.
\end{enumerate}\label{IPSDefinition}
\end{definition}
It was demonstrated by Goode and Wainwright \cite{Goode} that, in
order to ensure initial asymptotic isotropy, it is also necessary to
introduce a constraint on the cosmological fluid flow.
\begin{definition}[IPS fluid congruence]
With any unit timelike congruence $\mathbf{u}$ in $\mathcal{M}$ we
can associate a unit timelike congruence $\mathbf{\tilde{u}}$ in
$\tilde{\mathcal{M}}$ such that
\begin{eqnarray}
\mathbf{\tilde{u}} &=& \Omega\mathbf{u}\qquad \textrm{in }
\mathcal{M}\textrm{.}
\end{eqnarray}
\begin{itemize}
\item[a)] If we can choose $\mathbf{\tilde{u}}$ to be regular (at least $C^{3}$)
on an open neighbourhood of $T = 0$ in $\tilde{\mathcal{M}}$, we say
that $\mathbf{u}$ is regular at the IPS.
\item[b)] If, in addition, $\mathbf{\tilde{u}}$ is orthogonal to $T =
0$, we say that $\mathbf{u}$ is orthogonal to the IPS.
\end{itemize}\label{IPSFluidDefinition}
\end{definition}
In figure \ref{IPS} we present a pictorial interpretation of the
IPS.
\newpage
Below is given the analogous definition of an Isotropic Future
Singularity introduced by H\"ohn and Scott \cite{Scott}, followed by
the constraint on the fluid flow required to ensure final asymptotic
isotropy.
\begin{definition}[Isotropic Future Singularity (IFS)]
A space-time $(\mathcal{M},\mathbf{g})$ admits an Isotropic Future
Singularity if there exists a space-time $(\bar{\mathcal{M}},
\mathbf{\bar{g}})$, a smooth cosmic time function $\bar{T}$ defined
on $\bar{\mathcal{M}}$, and a conformal factor
$\bar{\Omega}(\bar{T})$ which satisfy
\begin{enumerate}
\item[i)] $\mathcal{M}$ is the open submanifold $\bar{T} < 0$,
\item[ii)] $\mathbf{g} = \bar{\Omega}^{2}(\bar{T})\mathbf{\bar{g}}$ on
$\mathcal{M}$, with $\mathbf{\bar{g}}$ regular (at least $C^{2}$ and
non-degenerate) on an open neighbourhood of $\bar{T} = 0$,
\item[iii)] $\bar{\Omega}(0) = 0$ and $\exists\; c > 0$ such that $\bar{\Omega} \in C^{0}[-c, 0]
\cap C^{2}[-c,0)$ and $\bar{\Omega}$ is positive on $[-c,0)$,
\item[iv)] $\bar{\lambda} \equiv \lim\limits_{\bar{T}\rightarrow 0^{-}}\bar{L}(\bar{T})$ exists, $\bar{\lambda} \neq
1$, where $\bar{L} \equiv
\frac{\bar{\Omega}''}{\bar{\Omega}}\left(\frac{\bar{\Omega}}{\bar{\Omega}'}\right)^{2}$
and a prime denotes differentiation with respect to $\bar{T}$.
\end{enumerate}\label{IFSDefinition}
\end{definition}
\begin{definition}[IFS fluid congruence]
With any unit timelike congruence $\mathbf{u}$ in $\mathcal{M}$ we
can associate a unit timelike congruence $\mathbf{\bar{u}}$ in
$\bar{\mathcal{M}}$ such that
\begin{eqnarray}
\mathbf{\bar{u}} &=& \bar{\Omega}\mathbf{u}\qquad \textrm{in }
\mathcal{M}\textrm{.}
\end{eqnarray}
\begin{itemize}
\item[a)] If we can choose $\mathbf{\bar{u}}$ to be regular (at least $C^{2}$)
on an open neighbourhood of $\bar{T} = 0$ in $\bar{\mathcal{M}}$, we
say that $\mathbf{u}$ is regular at the IFS.
\item[b)] If, in addition, $\mathbf{\bar{u}}$ is orthogonal to $\bar{T} =
0$, we say that $\mathbf{u}$ is orthogonal to the IFS.
\end{itemize}\label{IFSFluidDefinition}
\end{definition}
In figure \ref{IFS} we present a pictorial interpretation of the
IFS.
\\
Finally we give below the definition of a Future Isotropic Universe
introduced by H\"ohn and Scott \cite{Scott}. This definition covers
the further possibility for a conformal structure with an isotropic
future behaviour, which does not necessarily lead to a future
singularity; for example, some open FRW universes satisfy this
definition (see section \ref{ConformalFRW}).
\begin{definition}[Future Isotropic Universe
(FIU)] A space-time $(\mathcal{M},\mathbf{g})$ is said to be a
Future Isotropic Universe if there exists a space-time
$(\bar{\mathcal{M}}, \mathbf{\bar{g}})$, a smooth cosmic time
function $\bar{T}$ defined on $\bar{\mathcal{M}}$, and a conformal
factor $\bar{\Omega}(\bar{T})$ which satisfy
\begin{enumerate}
\item[i)] $\lim\limits_{\bar{T}\rightarrow 0^{-}}\bar{\Omega}(\bar{T}) = +\infty$ and $\exists\; c > 0$
such that $\bar{\Omega} \in C^{2}[-c,0)$ and $\bar{\Omega}$ is
strictly monotonically increasing and positive on $[-c,0)$,
\item[ii)] $\bar{\lambda}$ as defined above exists, $\bar{\lambda} \neq
1,2$, and $\bar{L}$ is continuous on $[-c,0)$ and
\item[iii)] otherwise the conditions of definitions \ref{IFSDefinition}
and \ref{IFSFluidDefinition} are satisfied.
\end{enumerate}\label{FIUDefinition}
\end{definition}
\section{Conformal Structures for the Cases $\eta_{0} < 0$ and $0 < \eta_{0} <
1$\label{ConformalFRW}} We recall the expression, given in section
\ref{CosmologicalEvents}, for the scale factor as a generalised
Puiseux series expansion around the time, $t_{0}$, when a
cosmological event occurs:
\begin{eqnarray}
a(t) &=& c_{0}|t-t_{0}|^{\eta_{0}} + c_{1}|t-t_{0}|^{\eta_{1}} +
c_{2}|t-t_{0}|^{\eta_{2}} +\cdots\qquad c_{0}>0\textrm{.}
\end{eqnarray}
Define $\tau = t-t_{0}$, so that $\tau \rightarrow 0^{+}$ at an
initial singularity and $\tau \rightarrow 0^{-}$ at a final
cosmological event. This means that the scale factor can be
expressed as
\begin{eqnarray}
a(\tau) &=& c_{0}|\tau|^{\eta_{0}} + c_{1}|\tau|^{\eta_{1}} +
c_{2}|\tau|^{\eta_{2}} + \cdots\\
&=& c_{0}|\tau|^{\eta_{0}}\left(1 +
\frac{c_{1}}{c_{0}}|\tau|^{\eta_{1} - \eta_{0}} +
\frac{c_{2}}{c_{0}}|\tau|^{\eta_{2}-\eta_{0}}+\cdots
\right)\textrm{,}\qquad\eta_{1}-\eta_{0} > 0
\end{eqnarray}
The metric of the FRW space-time $(\mathcal{M},g)$ is given, in the
coordinates $(\tau, r, \theta, \phi)$, by
\begin{eqnarray}
\fl\rmd s^{2} &=& -\rmd\tau^{2} + a^{2}(\tau)\left(\frac{\rmd
r^{2}}{1-kr^{2}} + r^{2}\left(\rmd\theta^{2} +
\sin^{2}\theta\rmd\phi^{2} \right) \right)\\
\fl&=&-\rmd\tau^{2} + c_{0}^{\phantom{0}2}\tau^{2\eta_{0}}\left(1 +
2\frac{c_{1}}{c_{0}}|\tau|^{\eta_{1}-\eta_{0}} +
\cdots\right)\rmd\sigma^{2}\textrm{,}\\
\fl\textrm{where }\rmd\sigma^{2} &=& \frac{\rmd r^{2}}{1-kr^{2}} +
r^{2}\left(\rmd\theta^{2} + \sin^{2}\theta\rmd\phi^{2}
\right)\textrm{.}\nonumber
\end{eqnarray}
Apart from the standard coordinate singularities, due to the use of
spherical polar coordinates, the metric has the following behaviour:
\begin{itemize}
\item[(i)] if $\eta_{0} < 0$ then there is a metric singularity at
$\tau = 0$, due to certain metric components becoming infinite,\\
\item[(ii)] if $\eta_{0} > 0$ then the metric becomes degenerate at
$\tau = 0$,\\
\item[(iii)] if $\eta_{0} = 0$ the metric may be regular at $\tau =
0$ or singular if any of the time derivatives with respect to $\tau$
diverge at $\tau = 0$.
\end{itemize}
In light of these facts, and using the previous section as
motivation, we will now investigate how these three cases can be
accommodated using the conformal frameworks introduced by H\"ohn and
Scott.
\subsection{The Case $\eta_{0} < 0$}\label{FRWFIU}
For an initial cosmological event define a new time coordinate $T$
by $T = \tau^{\alpha}$, $\alpha > 0$. So $T\rightarrow 0^{+}$ as
$\tau\rightarrow 0^{+}$. Define the manifold $\tilde{\mathcal{M}}$
to have the coordinate patch $(T,r,\theta,\phi)$, so that
$\mathcal{M}$ is the open submanifold $T > 0$ of
$\tilde{\mathcal{M}}$.\\

For a final cosmological event define another new time coordinate
$\bar{T}$ by $(-\bar{T}) = (-\tau)^{\alpha}$, $\alpha > 0$. So
$\bar{T}\rightarrow 0^{-}$ as $\tau\rightarrow 0^{-}$. Now define
the manifold $\bar{\mathcal{M}}$ to have the coordinate patch
$(\bar{T},r,\theta,\phi)$, so that $\mathcal{M}$ is the open
submanifold $\bar{T} < 0$ of $\bar{\mathcal{M}}$.\\

The metrics for the initial cosmological event and the final
cosmological event can now be expressed as, respectively,
\begin{eqnarray}
\fl\rmd s^{2} &=& -\frac{1}{\alpha^{2}}T^{\frac{2}{\alpha} - 2}\rmd
T^{2} + c_{0}^{\phantom{0}2}T^{\frac{2\eta_{0}}{\alpha}}\left(1 +
2\frac{c_{1}}{c_{0}}T^{\frac{\eta_{1} - \eta_{0}}{\alpha}} + \cdots
\right)\rmd\sigma^{2}
\end{eqnarray}
and
\begin{eqnarray}
\fl\rmd s^{2} &=& -\frac{1}{\alpha^{2}}(-\bar{T})^{\frac{2}{\alpha}
- 2}\rmd\bar{T}^{2} +
c_{0}^{\phantom{0}2}(-\bar{T})^{\frac{2\eta_{0}}{\alpha}}\left(1 +
2\frac{c_{1}}{c_{0}}(-\bar{T})^{\frac{\eta_{1} - \eta_{0}}{\alpha}}
+ \cdots \right)\rmd\sigma^{2}\textrm{.}
\end{eqnarray}
If we set $\alpha = 1-\eta_{0}$ (so that the range of $\alpha$ is
$\alpha > 1$) then the following conformal factors can be extracted,
respectively, from the metrics:
\begin{eqnarray}
\Omega(T) &=& T^{\frac{\eta_{0}}{1-\eta_{0}}}\textrm{,}\\
\bar{\Omega}(\bar{T}) &=&
(-\bar{T})^{\frac{\eta_{0}}{1-\eta_{0}}}\textrm{.}
\end{eqnarray}
Since $\eta_{0} < 0$ the exponent in each of these conformal factors
will be restricted to the range
\begin{eqnarray}
-1 < \frac{\eta_{0}}{1-\eta_{0}} < 0\textrm{.}
\end{eqnarray}
Using this, it can be seen that
\begin{itemize}
\item[(i)]$\lim\limits_{T\rightarrow 0^{+}}\Omega(T) =
+\infty$ and $\exists\; c > 0$ such that $\Omega \in C^{2}(0,c]$ and
$\Omega$ is strictly monotonically decreasing
and positive on $(0,c]$,\\
\item[(ii)]$\lim\limits_{\bar{T}\rightarrow 0^{-}}\bar{\Omega}(\bar{T}) =
+\infty$ and $\exists\; c > 0$ such that $\bar{\Omega} \in
C^{2}[-c,0)$ and $\bar{\Omega}$ is strictly monotonically increasing
and positive on $[-c,0)$.
\end{itemize}
We now list the first two derivatives of the conformal factors as
well as $L$ and $\bar{L}$:
\begin{eqnarray*}
\Omega' = \frac{\eta_{0}}{1-\eta_{0}}T^{\frac{2\eta_{0} -
1}{1-\eta_{0}}}&\qquad&\bar{\Omega}' =
-\frac{\eta_{0}}{1-\eta_{0}}(-\bar{T})^{\frac{2\eta_{0} -
1}{1-\eta_{0}}}\\
\Omega'' = \frac{\eta_{0}(2\eta_{0} -
1)}{(1-\eta_{0})^{2}}T^{\frac{3\eta_{0}-2}{1-\eta_{0}}}&\qquad&\bar{\Omega}''
= \frac{\eta_{0}(2\eta_{0} -
1)}{(1-\eta_{0})^{2}}(-\bar{T})^{\frac{3\eta_{0}-2}{1-\eta_{0}}}\\
L =
\frac{\Omega''\Omega}{\Omega'^{2}}=2-\frac{1}{\eta_{0}}&\qquad&\bar{L}
= \frac{\bar{\Omega}''\bar{\Omega}}{\bar{\Omega'^{2}}}=
2-\frac{1}{\eta_{0}}\textrm{.}
\end{eqnarray*}
We note that $L$ and $\bar{L}$ are constant functions and so are
continuous on $(0,c]$ and $[-c,0)$ respectively. Also $\lambda
\equiv \lim\limits_{T\rightarrow 0^{+}}L(T) = 2-\frac{1}{\eta_{0}}$
and $\bar{\lambda}\equiv \lim\limits_{\bar{T}\rightarrow
0^{-}}\bar{L}(\bar{T}) = 2-\frac{1}{\eta_{0}}$ and since $2 < 2 -
\frac{1}{\eta_{0}}$ it is clear that $\lambda$, $\bar{\lambda}\neq
1$, $2$.\\

The conformally related unphysical metrics, $\tilde{g}$ and
$\bar{g}$, are given by
\begin{eqnarray}
\label{TildeMetric}\rmd\tilde{s}^{2} &=&
-\frac{1}{(1-\eta_{0})^{2}}\rmd T^{2} + c_{0}^{\phantom{0}2}\left(1
+ 2\frac{c_{1}}{c_{0}}T^{\frac{\eta_{1}
- \eta_{0}}{1-\eta_{0}}} + \cdots \right)\rmd\sigma^{2}\textrm{,}\\
\label{BarMatric}\rmd\bar{s}^{2} &=&
-\frac{1}{(1-\eta_{0})^{2}}\rmd\bar{T}^{2} +
c_{0}^{\phantom{0}2}\left(1 +
2\frac{c_{1}}{c_{0}}(-\bar{T})^{\frac{\eta_{1} -
\eta_{0}}{1-\eta_{0}}} + \cdots \right)\rmd\sigma^{2}\textrm{,}
\end{eqnarray}
from which it is clear that the coordinates $T$ and $\bar{T}$ are
smooth cosmic time functions on $\tilde{\mathcal{M}}$ and
$\bar{\mathcal{M}}$ respectively. Since
$\frac{\eta_{1}-\eta_{0}}{1-\eta_{0}}
> 0$, the metrics $\tilde{g}$ and $\bar{g}$ are at least $C^{0}$
and non-degenerate on open neighbourhoods of $T=0$ and $\bar{T} = 0$
respectively. The precise differentiability of the metrics fall into
three classes:
\begin{itemize}
\item[(i)] If $\eta_{1} \geq 2-\eta_{0} > 2$, then $\tilde{g}$ and
$\bar{g}$ are at least $C^{2}$
\item[(ii)] If $1 \leq \eta_{1} < 2-\eta_{0}$, then $\tilde{g}$ and
$\bar{g}$ are $C^{1}$ (for $\eta_{1} = 1$, at least $C^{1}$)
\item[(iii)] If $\eta_{0} < \eta_{1} < 1$, then $\tilde{g}$ and $\bar{g}$ are
$C^{0}$.
\end{itemize}
The cosmological fluid flow $u$ given by
\begin{eqnarray}
u &=&
(1-\eta_{0})T^{-\frac{\eta_{0}}{1-\eta_{0}}}\frac{\partial}{\partial
T}\qquad\textrm{for an initial cosmological event}\\
u &=&
(1-\eta_{0})(-\bar{T})^{-\frac{\eta_{0}}{1-\eta_{0}}}\frac{\partial}{\partial
\bar{T}}\qquad\textrm{for a final cosmological event}
\end{eqnarray}
is a unit timelike congruence in $\mathcal{M}$. The corresponding
unit timelike congruences $\tilde{u}$ in $\tilde{\mathcal{M}}$ and
$\bar{u}$ in $\bar{\mathcal{M}}$ are
\begin{eqnarray}
\tilde{u} &=& \Omega u = (1-\eta_{0})\frac{\partial}{\partial T}\\
\bar{u} &=& \bar{\Omega} u = (1-\eta_{0})\frac{\partial}{\partial
\bar{T}}\textrm{.}
\end{eqnarray}
It is clear that we can choose $\tilde{u}$ and $\bar{u}$ to be
regular (at least $C^{2}$) on an open neighbourhood of $T = 0$ in
$\tilde{\mathcal{M}}$ and an open neighbourhood of $\bar{T} = 0$ in
$\bar{\mathcal{M}}$ respectively. Additionally $\tilde{u}$ and
$\bar{u}$ are orthogonal to $T=0$ and $\bar{T} = 0$ respectively.\\

We have examined the FRW space-times for a general expansion of the
scale factor for the case $\eta_{0} < 0$ and have seen that these
metrics all exhibit metric singularities. The scale factor has the
behaviour $a(t)\rightarrow +\infty$ as $t\rightarrow t_{0}$ so these
space-times all possess Big Rips \cite{Visser} in either their past
or future. In the conformal framework given by H\"ohn and Scott the
physical metrics represent a Past Isotropic Universe or a Future
Isotropic Universe. At the cosmological event in question, they are
conformally related to a regular metric which exists on a larger
manifold; these conformally related regular metrics have varying
degrees of differentiability. The conformal factor diverges at the
cosmological event and the unphysical fluid flow is regular on an
open neighbourhood of the event and orthogonal to it.
\subsection{The Case $0 < \eta_{0} < 1$}\label{FRWIPS}
Having examined the case $\eta_{0} < 0$ in detail we now turn our
attention to the case $\eta_{0} > 0$. The investigation of the case
$0 < \eta_{0} < 1$ is very similar to the case $\eta_{0} < 0$ and is
presented first. We only present the analysis for a final
cosmological event; the initial cosmological event follows logically
as seen in the case $\eta_{0}<0$.\\

Define a new time coordinate $\bar{T}$ by $(-\bar{T}) =
(-\tau)^{\alpha}$, $\alpha
> 0$. So $\bar{T}\rightarrow 0^{-}$ as $\tau\rightarrow 0^{-}$. Now
define the manifold $\bar{\mathcal{M}}$ to have the coordinate patch
$(\bar{T},r,\theta,\phi)$ so that $\mathcal{M}$ is the open
submanifold $\bar{T} < 0$ of $\bar{\mathcal{M}}$.\\

The metric for the final cosmological event can be reexpressed as
\begin{eqnarray}
\fl\rmd s^{2} &=& -\frac{1}{\alpha^{2}}(-\bar{T})^{\frac{2}{\alpha}
- 2}\rmd\bar{T}^{2} +
c_{0}^{\phantom{0}2}(-\bar{T})^{\frac{2\eta_{0}}{\alpha}}\left(1 +
2\frac{c_{1}}{c_{0}}(-\bar{T})^{\frac{\eta_{1} - \eta_{0}}{\alpha}}
+ \cdots \right)\rmd\sigma^{2}
\end{eqnarray}
If we set $\alpha = 1-\eta_{0}$ (so that the range of $\alpha$ is $0
< \alpha < 1$) then the following conformal factor can be extracted
from the metric:
\begin{eqnarray}
\bar{\Omega}(\bar{T}) &=&
(-\bar{T})^{\frac{\eta_{0}}{1-\eta_{0}}}\textrm{.}
\end{eqnarray}
Since $0 < \eta_{0} < 1$ the exponent will always be greater than
zero. Using this, it can be seen that
\begin{itemize}
\item[(i)] $\lim\limits_{\bar{T}\rightarrow 0^{-}}\bar{\Omega}(\bar{T}) =
0$ and $\exists\; c > 0$ such that $\bar{\Omega} \in C^{0}[-c,0]\cap
C^{2}[-c,0)$ and $\bar{\Omega}$ is strictly monotonically decreasing
and positive on $[-c,0)$.
\end{itemize}
We note that $\bar{L}$ is the same as for the PIU/FIU seen
previously
\begin{eqnarray*}
\bar{L} &=& \frac{\bar{\Omega}''\bar{\Omega}}{\bar{\Omega}'^{2}}=
2-\frac{1}{\eta_{0}}\textrm{.}\\
\end{eqnarray*}
Also $\bar{\lambda}\equiv \lim\limits_{\bar{T}\rightarrow
0^{-}}\bar{L}(\bar{T}) = 2 - \frac{1}{\eta_{0}} < 1$. The
conformally related metric $\bar{g}$ is the same as the metric for
the FIU
\begin{eqnarray}
\rmd\bar{s}^{2} &=& -\frac{1}{(1-\eta_{0})^{2}}\rmd\bar{T}^{2} +
c_{0}^{\phantom{0}2}\left(1 +
2\frac{c_{1}}{c_{0}}(-\bar{T})^{\frac{\eta_{1} -
\eta_{0}}{1-\eta_{0}}} + \cdots \right)\rmd\sigma^{2}\textrm{,}
\end{eqnarray}
from which it is clear that the coordinate $\bar{T}$ is a smooth
cosmic time function on $\bar{\mathcal{M}}$. Since
$\frac{\eta_{1}-\eta_{0}}{1-\eta_{0}}
> 0$, the metric $\bar{g}$ is at least $C^{0}$ and non-degenerate on an
open neighbourhood of $\bar{T} = 0$. The precise differentiability
of the metric again falls into three classes:
\begin{itemize}
\item[(i)] If $\eta_{1} \geq 2-\eta_{0} \in (1,2)$, then
$\bar{g}$ is at least $C^{2}$
\item[(ii)] If $1 \leq \eta_{1} < 2-\eta_{0} \in (1,2)$, then
$\bar{g}$ is $C^{1}$ (for $\eta_{1} = 1$, at least $C^{1}$)
\item[(iii)] If $\eta_{0} < \eta_{1} < 1$, then $\bar{g}$ is
$C^{0}$.
\end{itemize}
The cosmological fluid flow $u$ given by
\begin{eqnarray}
u &=&
(1-\eta_{0})(-\bar{T})^{-\frac{\eta_{0}}{1-\eta_{0}}}\frac{\partial}{\partial
\bar{T}}
\end{eqnarray}
is a unit timelike congruence in $\mathcal{M}$. The corresponding
unit timelike congruence $\bar{u}$ in $\mathcal{M}$ is
\begin{eqnarray}
\bar{u} &=& \bar{\Omega} u = (1-\eta_{0})\frac{\partial}{\partial
\bar{T}}\textrm{.}
\end{eqnarray}
Again, it is clear that we can choose $\bar{u}$ to be regular (at
least $C^{2}$) on an open neighbourhood of $\bar{T} = 0$ in
$\bar{\mathcal{M}}$ and that $\bar{u}$ is orthogonal to $\bar{T} =
0$.\\

We have examined the case $\eta_{0} \in (0,1)$ and have seen that
the metric exhibits a metric singularity. The scale factor has the
behaviour $a(t)\rightarrow 0$ as $t\rightarrow t_{0}$ so the metric
possesses a Big Bang if the singularity is in the past or a Big
Crunch if the singularity is in the future \cite{Visser}. According
to H\"ohn and Scott this metric represents an Isotropic Past
Singularity or an Isotropic Future Singularity respectively. At the
cosmological event in question, the metric is conformally related to
a regular metric which exists on a larger manifold; the conformally
related regular metric has varying degrees of differentiability. The
conformal factor converges to $0$ at the cosmological event and the
unphysical fluid flow is regular on an open neighbourhood of the
event and orthogonal to it.
\section{The Case $\eta_{0} \geq 1$}
Logically, the next case we would want to consider is for
$\eta_{0}\geq 1$. This case is substantially more involved than the
cases $\eta_{0} < 0$ and $0 < \eta_{0} < 1$ and requires a great
deal of care.\\

The first thing we demonstrate is that these solutions cannot
represent an IPS/IFS. We need to see, for the case of
$\eta_{0}\geq1$, if $\frac{\ddot{a}a}{(\dot{a})^{2}}\geq 0$. If this
is true then this class of FRW models cannot
admit an IPS/IFS (due to The FRW Result of Ericksson \cite{Ericksson}).\\

The scale factor and its derivatives are
\begin{eqnarray}
a(\tau) &=& \sum_{i=0} c_{i}\tau^{\eta_{i}}\\
\dot{a}(\tau) &=& \sum_{i=0} \eta_{i}c_{i}\tau^{\eta_{i}-1}\\
\ddot{a}(\tau) &=& \sum_{i=0} \eta_{i}\left(\eta_{i} -
1\right)c_{i}\tau^{\eta_{i}-2}
\end{eqnarray}
where $c_{0}\neq 0$, $\eta_{0} \neq 0$ and $\eta_{0} \neq 1$.\\

To leading order we see,
\begin{eqnarray}
\frac{\ddot{a}(\tau)a(\tau)}{\dot{a}^{2}(\tau)} &\approx&
\frac{c_{0}\eta_{0}(\eta_{0}-1)\tau^{\eta_{0}-2}\cdot
c_{0}\tau^{\eta_{0}}}{\left(c_{0}\eta_{0}\tau^{\eta_{0}-1}\right)^{2}}\\
&=& \frac{\eta_{0}-1}{\eta_{0}}\textrm{.}
\end{eqnarray}
So we see that
\begin{eqnarray}
\eta_{0} < 0 &\Longrightarrow& \frac{\ddot{a}a}{\dot{a}^{2}} > 1\\
0 < \eta_{0} < 1 &\Longrightarrow& \frac{\ddot{a}a}{\dot{a}^{2}} < 0\\
\eta_{0} = 1 &\Longrightarrow& \lim\limits_{\tau\to
0^{\pm}}\frac{\ddot{a}a}{\dot{a}^{2}} \to 0\\
\eta_{0} > 1 &\Longrightarrow& 0 < \frac{\ddot{a}a}{\dot{a}^{2}} < 1\textrm{.}
\end{eqnarray}

This demonstrates that, for $\eta_{0} \geq 1$ and for $\eta_{0} <
0$, the function $\frac{\ddot{a}a}{\dot{a}^{2}} > 0$ and as such
these type of FRW models cannot admit an IPS/IFS \cite{Ericksson} at
which the fluid flow is regular. We do know that for $\eta_{0} < 0$,
these FRW models obey the PIU/FIU definition as seen in section
\ref{FRWFIU}.\\

We seek to exclude the possibility that for $\eta_{0} \geq 1$, these
solutions admit a PIU/FIU; this is the direction of our analysis
now. This study is very similar to the FRW result of Ericksson
\cite{Ericksson}. We will need to prove a number
of preliminary results before we present that result and provide those first.\\

We introduce the following condition.
\begin{condition}
The spacetime $(\mathcal{M},g)$ is a $C^{3}$ solution of the
Einstein Field Equations with perfect fluid source and the unit
timelike fluid congruence $u$ is regular at a
PIU/FIU.\label{ConditionStar}
\end{condition}
This condition is used to consider perfect fluid sources (the FRW
solutions fall within this).

\begin{theorem}
If the spacetime $(\mathcal{M},g)$ satisfies condition
\ref{ConditionStar} then the unit timelike fluid congruence
$\bar{u}$ is orthogonal to the PIU/FIU.
\end{theorem}
\emph{Proof}\\

Similar proofs have previously been presented first by Scott
\cite{Scott2} and
subsequently Ericksson \cite{Ericksson}.\\

For a perfect fluid the Einstein Field Equations imply that the rank
one tensor representing the anisotropic parts of the Ricci tensor,
$\Sigma_{a}$, vanishes. Using this as motivation, we proceed as
follows
\begin{eqnarray}
\Sigma_{a} = 0\Longleftrightarrow \Sigma^{a} = 0\\
\Longrightarrow \Sigma^{a}\bar{T}_{,a} = 0
\end{eqnarray}
where
\begin{eqnarray}
\Sigma^{a} = \bar{\Omega}^{-3}\left(\bar{\Sigma}^{a} +
2\bar{h}^{ab}\left(\bar{\Omega}^{-1}\bar{\Omega}_{;bc} -
2\bar{\Omega}^{-2}\bar{\Omega}_{,b}\bar{\Omega}_{,c}
\right)\bar{u}^{c}\right)\textrm{.}
\end{eqnarray}
This means that we can write the equation $\Sigma^{a}\bar{T}_{,a} =
0$ as
\begin{eqnarray}
& &\bar{\Omega}^{-3}\left(\bar{\Sigma}^{a}\bar{T}_{,a} +
2\bar{h}^{ab}\bar{T}_{,a}\left(\bar{\Omega}^{-1}\bar{\Omega}_{;bc} -
2\bar{\Omega}^{-2}\bar{\Omega}_{,b}\bar{\Omega}_{,c}
\right)\bar{u}^{c}\right) = 0\textrm{.}\label{SigmaOne}
\end{eqnarray}
If the conformal factor is a function of cosmic time only,
$\bar{\Omega} = \bar{\Omega}(\bar{T})$, as is true for our analysis,
then
\begin{eqnarray}
\bar{\Omega}_{,a} &=& \bar{\Omega}'\bar{T}_{,a}\\
\bar{\Omega}_{;ab} &=& \bar{\Omega}'\left(\bar{T}_{;ab} +
\frac{\bar{\Omega}''}{\bar{\Omega}'}\bar{T}_{,a}\bar{T}_{,b}
\right)\textrm{.}
\end{eqnarray}
This simplifies equation \ref{SigmaOne} to
\begin{eqnarray}
& &\bar{\Omega}^{-3}\left(\bar{\Sigma}^{a}\bar{T}_{,a} +
2\bar{h}^{ab}\bar{T}_{,a}\frac{\bar{\Omega}'}{\bar{\Omega}}\left(\bar{T}_{;bc}
+ \left(\bar{L} -
2\right)\frac{\bar{\Omega}'}{\bar{\Omega}}\bar{T}_{,b}\bar{T}_{,c}
\right)\bar{u}^{c}\right) = 0
\end{eqnarray}
From the definition of a PIU/FIU we know that $\exists\; b
> 0$ such that $\bar{\Omega}$ is positive on $[-b,0)$, also
$\exists\; c\in (0,b]$ such that $\bar{\Omega}'\neq 0$ on $[-c,0)$.
So therefore
\begin{eqnarray}
\left(\frac{\bar{\Omega}}{\bar{\Omega}'}\right)^{2}\bar{\Sigma}^{a}\bar{T}_{,a}
+
2\frac{\bar{\Omega}}{\bar{\Omega}'}\bar{h}^{ab}\bar{T}_{,a}\bar{T}_{;bc}\bar{u}^{c}
+ 2\left(\bar{L} -
2\right)\bar{h}^{ab}\bar{T}_{,a}\bar{T}_{,b}\bar{T}_{,c}\bar{u}^{c}
= 0
\end{eqnarray}
is valid on $[-c,0)$. We also know that
$\bar{\Sigma}^{a}\bar{T}_{,a}$ is at least $C^{1}$,
$\bar{h}^{ab}\bar{T}_{,a}\bar{T}_{;bc}\bar{u}^{c}$ is at least
$C^{2}$ and
$\bar{h}^{ab}\bar{T}_{,a}\bar{T}_{,b}\bar{T}_{,c}\bar{u}^{c}$ is at
least $C^{3}$ on an open neighbourhood of $\bar{T} = 0$ in
$\bar{\mathcal{M}}$. We know that as $\bar{T}\to 0^{-}$,
$\frac{\bar{\Omega}}{\bar{\Omega}'}\to 0$ and therefore as
$\bar{T}\to 0$
\begin{eqnarray}
& &
\left(\frac{\bar{\Omega}}{\bar{\Omega}'}\right)^{2}\bar{\Sigma}^{a}\bar{T}_{,a}
+
2\frac{\bar{\Omega}}{\bar{\Omega}'}\bar{h}^{ab}\bar{T}_{,a}\bar{T}_{;bc}\bar{u}^{c}
\to 0\\
& & \Longrightarrow 2\left(\bar{L} -
2\right)\bar{h}^{ab}\bar{T}_{,a}\bar{T}_{,b}\bar{T}_{,c}\bar{u}^{c}
= 0\textrm{.}\label{SigmaOneA}
\end{eqnarray}
Now $\bar{\lambda}\neq 2$ and $\bar{u}^{c}\bar{T}_{,c}\neq 0$ on
$\bar{T}=0$ so therefore $\bar{h}^{ab}\bar{T}_{,a}\bar{T}_{,b} = 0$
on $\bar{T} = 0$ and therefore $\bar{u}$ is orthogonal to
$\bar{T}=0$. $\Box$

\begin{theorem}
If the spacetime $(\mathcal{M},g)$ satisfies condition
\ref{ConditionStar} and $1 < \bar{\lambda} < +\infty$ then there
exists a limiting $\gamma$-law equation of state $p = (\gamma-1)\mu$
as the PIU/FIU is approached, where $\gamma =
\frac{2}{3}(2-\bar{\lambda})$.
\end{theorem}
\emph{Proof}\\

For a perfect fluid source, the Einstein Field Equations reveal the
following \cite{Ericksson}
\begin{eqnarray}
A &=& \mu =
\bar{\Omega}^{-2}\Big[\left(\frac{\bar{\Omega}'}{\bar{\Omega}}\right)^{2}\left[3\left(\bar{u}^{a}\bar{T}_{,a}\right)^{2}
+ (1-2\bar{L})\bar{h}^{ab}\bar{T}_{,a}\bar{T}_{,b}\right] -
2\frac{\bar{\Omega}'}{\bar{\Omega}}\bar{h}^{ab}\bar{T}_{;ab}\nonumber\\
&+& \bar{A}\Big]\\
B &=& p =
\bar{\Omega}^{-2}\Big[\left(\frac{\bar{\Omega}'}{\bar{\Omega}}\right)^{2}\left[(1-2\bar{L})\left(\bar{u}^{a}\bar{T}_{,a}\right)^{2}
+ \frac{1}{3}(1+4\bar{L})\bar{h}^{ab}\bar{T}_{,a}\bar{T}_{,b}\right]\nonumber\\
&+&
\frac{2}{3}\frac{\bar{\Omega}'}{\bar{\Omega}}\left(\bar{h}^{ab}\bar{T}_{;ab}
- 3\bar{u}^{a}\bar{u}^{b}\bar{T}_{;ab}\right) + \bar{B}\Big]
\end{eqnarray}
Now we know that as $\bar{T}\to 0$, $\bar{\Omega}\to+\infty$,
$\bar{\Omega}'/\bar{\Omega}\to+\infty$ and the unphysical metric is
$C^{2}$, so we can approximate the equations above as
\begin{eqnarray}
\mu &\approx&
3\bar{M}^{2}\left(\bar{u}^{a}\bar{T}_{,a}\right)^{2}\\
p &\approx&
\bar{M}^{2}\left(1-2\bar{\lambda}\right)\left(\bar{u}^{a}\bar{T}_{,a}\right)^{2}
\end{eqnarray}
where $\bar{M} = \bar{\Omega}'/\bar{\Omega}^{2}$. The difficulty
here, unlike for the case of the IPS/IFS, is that $\bar{M}$ does not
necessarily diverge to $+\infty$; it will have different behaviour
for $\bar{\Omega}(0) = +\infty$ as compared to when $\bar{\Omega}(0)
= 0$. The behaviour of $\bar{M}$ has previously been determined
\cite{Scott} and as such we simply exploit the results below.\\

For $1<\bar{\lambda}<2$, $\lim\limits_{\bar{T}\to
0^{-}}\bar{M}\in\mathbb{R}^{+}\cup\{0\}$, while for $\bar{\lambda}
> 2$ $\lim\limits_{\bar{T}\to
0^{-}}\bar{M}\in\mathbb{R}^{+}\cup\{+\infty\}$. The case
$\bar{\lambda} = 2$ is eliminated by the definition of a PIU/FIU.\\

The asymptotic behaviour for $\mu$ and $p$ as $\bar{T}\to 0^{-}$ is
as follows
\begin{eqnarray}
\mu &\approx&
3\bar{M}^{2}\left(\bar{u}^{a}\bar{T}_{,a}\right)^{2}\to \alpha \in\mathbb{R}^{+}\cup\{0\}\qquad 1<\bar{\lambda}<2\\
\mu &\approx&
3\bar{M}^{2}\left(\bar{u}^{a}\bar{T}_{,a}\right)^{2}\to \beta \in\mathbb{R}^{+}\cup\{+\infty\}\qquad \bar{\lambda}>2\\
p &\approx&
\bar{M}^{2}\left(1-2\bar{\lambda}\right)\left(\bar{u}^{a}\bar{T}_{,a}\right)^{2}\to
\gamma \in \mathbb{R}^{-}\cup \{0\}\qquad
1<\bar{\lambda}<2\\
p &\approx&
\bar{M}^{2}\left(1-2\bar{\lambda}\right)\left(\bar{u}^{a}\bar{T}_{,a}\right)^{2}\to
\delta \in \mathbb{R}^{-}\cup \{-\infty\}\qquad \bar{\lambda} > 2
\end{eqnarray}
It follows that  for $1 < \bar{\lambda} < +\infty$
\begin{eqnarray}
p &\approx& \frac{1}{3}\left(1-2\bar{\lambda}\right)\mu
\end{eqnarray}
and so there exists a limiting $\gamma$-law equation of state
\begin{eqnarray}
p &=& \left(\gamma-1\right)\mu
\end{eqnarray}
where $\gamma = \frac{2}{3}(2-\bar{\lambda})$. $\Box$\\

We note that a solution is asymptotic dust if and only if
$\bar{M}\to 0$ as $\bar{T}\to 0^{-}$. This can only
occur for $1 < \bar{\lambda} < 2$.\\

\begin{theorem}
If the spacetime $(\mathcal{M},g)$ satisfies condition
\ref{ConditionStar} and $\bar{\lambda} = +\infty$ then there exists
a limiting equation of state $p\approx -\frac{2}{3}\bar{L}\mu$ as
the PIU/FIU is approached.
\end{theorem}
\emph{Proof}\\

Again we use the following equations from the Einstein Field
Equations
\begin{eqnarray}
A &=& \mu =
\bar{\Omega}^{-2}\Big[\left(\frac{\bar{\Omega}'}{\bar{\Omega}}\right)^{2}\left[3\left(\bar{u}^{a}\bar{T}_{,a}\right)^{2}
+ (1-2\bar{L})\bar{h}^{ab}\bar{T}_{,a}\bar{T}_{,b}\right] -
2\frac{\bar{\Omega}'}{\bar{\Omega}}\bar{h}^{ab}\bar{T}_{;ab}\nonumber\\
&+& \bar{A}\Big]\\
B &=& p =
\bar{\Omega}^{-2}\Big[\left(\frac{\bar{\Omega}'}{\bar{\Omega}}\right)^{2}\left[(1-2\bar{L})\left(\bar{u}^{a}\bar{T}_{,a}\right)^{2}
+ \frac{1}{3}(1+4\bar{L})\bar{h}^{ab}\bar{T}_{,a}\bar{T}_{,b}\right]\nonumber\\
&+&
\frac{2}{3}\frac{\bar{\Omega}'}{\bar{\Omega}}\left(\bar{h}^{ab}\bar{T}_{;ab}
- 3\bar{u}^{a}\bar{u}^{b}\bar{T}_{;ab}\right) + \bar{B}\Big]
\end{eqnarray}
Now we know that as $\bar{T}\to 0^{-}$ both $\bar{\Omega}$ and
$\bar{\Omega}'/\bar{\Omega}$ diverge and the unphysical metric is
$C^{2}$. We also know from equation \ref{SigmaOneA} that
$\bar{h}^{ab}\bar{T}_{,a}\bar{T}_{,b}\to0$ as $\bar{T}\to 0^{-}$.
Thus we can approximate the above equations as
\begin{eqnarray}
\mu &\approx&
3\bar{M}^{2}\left(\bar{u}^{a}\bar{T}_{,a}\right)^{2}\\
p &\approx&
-2\bar{L}\bar{M}^{2}\left(\bar{u}^{a}\bar{T}_{,a}\right)^{2}
\end{eqnarray}
where $\bar{M} = \bar{\Omega}'/\bar{\Omega}^{2}$.\\

For $\bar{\lambda} = +\infty$,
$\lim\limits_{\bar{T}\to0^{-}}\bar{M}$(exists)$=
\epsilon \in \mathbb{R}^{+}\cup\{+\infty\}$.\\

This means that, as $\bar{T}\to0^{-}$,
\begin{eqnarray}
\mu &\approx&
3\bar{M}^{2}\left(\bar{u}^{a}\bar{T}_{,a}\right)^{2}\to \zeta \in\mathbb{R}^{+}\cup\{+\infty\}\\
p &\approx&
-2\bar{L}\bar{M}^{2}\left(\bar{u}^{a}\bar{T}_{,a}\right)^{2}\to
-\infty
\end{eqnarray}
The limiting equation of state for $\bar{\lambda} = +\infty$ is
therefore
\begin{eqnarray}
p &\approx& -\frac{2}{3}\bar{L}\mu\qquad\Box
\end{eqnarray}
The following lemma is the equivalent of lemma 6.1.2 from
\cite{Ericksson} but for a PIU/FIU.
\begin{lemma}\label{ErickssonLemma}
Define a function $G(t)$ on $(0,t_{1}]$ by
\begin{eqnarray}
G(t)&\equiv& \int_{t}^{t_{1}}a(u)du
\end{eqnarray}
An FRW model satisfies all but one of the conditions of a PIU/FIU if
$G(t)\to \eta \in \mathbb{R}^{+}$ as $t\to 0$.
\end{lemma}
\emph{Proof}\\

The only condition that this does not immediately satisfy is the condition concerning $\bar{\lambda}$.\\

We can see that the function $G(t)$ is a positive and strictly
monotonically decreasing function on $(0,t_{1}]$ decreasing to zero
at $t_{1}$. This guarantees that $\mathop {\lim}\limits_{T \to
0}a(t)$ always exists, so that the condition $a(t)\to
\eta\in\mathbb{R}^{+}$ as $t\to 0$ only eliminates those
FRW models for which $G(t)\to+\infty$ as $t\to 0$.\\

If we assume that $G(t)\to \eta\in\mathbb{R}^{+}$ as $t\to 0$,
define $T$ on $(0,t_{1}]$ by
\begin{eqnarray}
T(t)&\equiv&
\lim_{\epsilon\to0}\int_{\epsilon}^{t}\frac{1}{a(u)}du
\end{eqnarray}

We see that $T(t)$ is well defined and positive, strictly
monotonically increasing on $(0,t_{1}]$ and that $T(t)\to 0$
as $t\to 0$.\\

The rest of the proof is essentially identical to that seen in
Ericksson \cite{Ericksson}. $\Box$

\begin{corollary}\label{ErickssonCorollary}
If an FRW model satisfies the conditions of Lemma
\ref{ErickssonLemma} then the fluid flow is regular at the PIU/FIU.
\end{corollary}

\emph{Proof}\\

The proof is the same as in Ericksson \cite{Ericksson} and as such
we direct the
reader there. $\Box$\\

If the spacetime admits a PIU/FIU then H\"ohn and Scott \cite{Scott}
has shown that $\bar{\lambda}> 1$ and $\bar{\lambda}\neq 1,\;2$;
this fact will be
key to determining the behaviour of a PIU/FIU.\\

For convenience, we now concisely present various properties of the
PIU/FIU. These results are due to \cite{Scott} or the previous
section.\\

For a PIU/FIU structure

\begin{eqnarray}\label{Variables}
\begin{array}{l}
\bar{\lambda} = \lim\limits_{\bar{T}\to0}\bar{L}
=\lim\limits_{\bar{T}\to0}
\frac{\bar{\Omega}''\bar{\Omega}}{\bar{\Omega}'^{2}}\\
\bar{M} = \frac{\bar{\Omega}'}{\bar{\Omega}^{2}}\\
a\equiv\bar{\Omega}\\
p \approx
\bar{M}^{2}\left(1-2\bar{\lambda}\right)\left(\bar{u}^{a}\bar{T}_{,a}\right)^{2}\\
\mu \approx 3\bar{M}^{2}\left(\bar{u}^{a}\bar{T}_{,a}\right)^{2}
\end{array}
\end{eqnarray}
In the limit that $\bar{T}\to 0$, the variables in equation
\ref{Variables} adopt the following asymptotic form
\begin{table}[h!]
\begin{tabular}{|c|c|c|c|c|}
\hline \hline $\bar{\lambda}$ & $\bar{M}$ & $a$ & $p$ & $\mu$\\
\hline $1 < \bar{\lambda} < 2$ & $\beta\in \mathbb{R}^{+}\cup\{0\}$ & $+\infty$ & $\gamma\in\mathbb{R}^{-}\cup\{0\}$& $\delta \in \mathbb{R}^{+}\cup\{0\}$\\
\hline $2 < \bar{\lambda} < +\infty$ & $\alpha\in\mathbb{R}^{+}\cup\{+\infty\}$ & $+\infty$ & $\epsilon\in\mathbb{R}^{-}\cup\{-\infty\} $ &$\phi \in \mathbb{R}^{+}\cup\{+\infty\}$\\
\hline $\bar{\lambda} = +\infty$ & $\alpha\in\mathbb{R}^{+}\cup\{+\infty\}$ & $+\infty$ & $-\infty$ &$\phi \in \mathbb{R}^{+}\cup\{+\infty\}$\\
\hline\hline
\end{tabular}
\end{table}\clearpage
 Using all our preliminary results, we now present the
following theorem that provides the final piece in puzzle for the
analysis of the FRW solution case $\eta_{0} \geq 1$.
\begin{theorem}[The FRW PIU/FIU Result]
An FRW model admits a PIU/FIU with $1 < \bar{\lambda} < +\infty$ at
which the fluid flow is regular if and only if $\mathop
{\lim}\limits_{T \to 0}\frac{\ddot{a}a}{\dot{a}^{2}}$ exists and is
greater than zero and $\mathop {\lim}\limits_{t \to 0}a(t)\to
+\infty$
\end{theorem}
\emph{Proof}\\

Assuming the existence of a PIU/FIU first, we seek to prove $\mathop
{\lim}\limits_{T \to 0}\frac{\ddot{a}a}{\dot{a}^{2}}$.\\

The equation of state is
\begin{eqnarray}
p &=& \left(\gamma-1\right)\mu\left[1+ f(t)\right]\qquad 1 < \lambda < +\infty\\
\end{eqnarray}
where $\gamma = \frac{2}{3}(2-\lambda)$ and $f(t) = o(1)$. We see
that if we use the conservation equations for a perfect fluid source
\cite{Ellis} we get
\begin{eqnarray}
\dot{\mu} &=& -3(\mu + p)\frac{\dot{a}}{a}\\
&=&-3\left(\mu + \left(\gamma-1\right)\mu\left[1+ o(1)\right]\right)\frac{\dot{a}}{a}\\
&=& -3\gamma\mu\frac{\dot{a}}{a}\left[1+ o(1)\right]
\end{eqnarray}
which means we can write the same as Ericksson \cite{Ericksson}
\begin{eqnarray}
\frac{\dot{\mu}}{\mu} &=& -3\gamma\left[1+ o(1)\right]\frac{\dot{a}}{a}\\
\ln\mu|_{t}^{t_{1}} &=& -3\gamma\left[1+ o(1)\right]\ln a|_{t}^{t_{1}} \\
\ln\mu &=& -3\gamma\left[1+ o(1)\right]\ln a\\
 \mu a^{2} &=& a^{2-3\gamma}
\end{eqnarray}
We need to examine what happens with $\gamma =
\frac{2}{3}(2-\lambda)$ as we vary $\bar{\lambda}$
\begin{eqnarray}
1 < \bar{\lambda} < 2 &\Longrightarrow& 0 < \gamma < \frac{2}{3}\\
\bar{\lambda} = 2 &\Longrightarrow&  \gamma = 0\\
\bar{\lambda} > 2 &\Longrightarrow& \gamma <0
\end{eqnarray}
So it means that we can say, because $a\to +\infty$ as $T\to 0$ for
a PIU/FIU
\begin{eqnarray}
\mathop {\lim}\limits_{\bar{T} \to 0}\mu a^{2} \to +\infty
\end{eqnarray}
This even includes the finite values that $\mu$ may take because $\mu$ is nonnegative.\\

 This means that due to the Einstein field equation,
\begin{eqnarray}
3\dot{a}^{2} &=& \mu a^{2} - 3K
\end{eqnarray}
it means $\mathop {\lim}\limits_{\bar{T} \to 0}\dot{a}^{2}\to
+\infty$.\\

Now we can use other field equations and evaluate $\ddot{a}/a$.
\begin{eqnarray}
\frac{\ddot{a}}{a} &=& -\frac{1}{6}\left(\mu + 3p\right)\\
&=& -\frac{1}{6}\left(\mu + 3\left(\gamma-1\right)\mu\left[1+ o(1)\right]\right)\\
&=& -\frac{\mu}{6}\left(3\gamma-2 + o(1)\right)
\end{eqnarray}
Now we can multiply this by $\frac{a^{2}}{\dot{a}^{2}}$
\begin{eqnarray}
\frac{\ddot{a}a}{\dot{a}^{2}} &=& -\frac{\mu
a^{2}}{6\dot{a}^{2}}\left(3\gamma-2\right)\left[1+ o(1)\right]\\
&=&-\frac{1}{6}\left(3\gamma-2\right) \left(3 +
\frac{3K}{\dot{a}^{2}} \right)\left[1+ o(1)\right]
\end{eqnarray}
As $\dot{a}^{2}\to +\infty$, it means
\begin{eqnarray}
\frac{\ddot{a}a}{\dot{a}^{2}}
&\to&-\frac{1}{2}\left(3\gamma-2\right)
\end{eqnarray}
For a PIU/FIU $\gamma < \frac{2}{3}$ and as such this value is
always going to be positive. Hence for a PIU/FIU
$\frac{\ddot{a}a}{\dot{a}^{2}} > 0$.\\

We will now prove the converse, again following the lead of
\cite{Ericksson}. Assume that the FRW spacetime has $\mathop
{\lim}\limits_{\bar{T} \to 0} \frac{\ddot{a}a}{\dot{a}^{2}} > 0$
exists. Let the limit of $\frac{\ddot{a}a}{\dot{a}^{2}}$ be written
as $\bar{\lambda}-1$. As $\bar{\lambda} > 1$ means this value will
always be positive.\\

We have assumed $\mathop {\lim}\limits_{t \to
0}\frac{\ddot{a}a}{\dot{a}^{2}} = \bar{\lambda} - 1 > 0$ and as such
\begin{eqnarray}
& &a = \left(\bar{\lambda} -
1\right)\frac{\dot{a}^{2}}{\ddot{a}}\left[1+ o(1)\right]\\
& & \int_{t}^{t_{1}} a(u)du =  \left(\bar{\lambda} -
1\right)\left(\int_{t}^{t_{0}}\frac{\dot{a}^{2}}{\ddot{a}}du\right)\left[1+
o(1)\right]
\end{eqnarray}
We need to find out if
$\int_{t}^{t_{0}}\frac{\dot{a}^{2}}{\ddot{a}}du$ is positive because
if it is then we know that $\int_{t}^{t_{1}} a(u)du$ will be
positive and we will have three of four of the conditions for a
PIU/FIU satisfied.\\

We know that $\dot{a}^{2}\to+\infty$ which means
$\ln\dot{a}^{2}\to+\infty = 2\ln\dot{a}$ as such there will be an
interval, $(0,t_{0}]$ such that $\dot{a}>0$. If we rewrite this as
\begin{eqnarray}
2\ln\dot{a} &=& 2\int\frac{\ddot{a}}{\dot{a}}du\to+\infty
\end{eqnarray}
So we know that because $\dot{a}>0$ it will mean that $\ddot{a}>0$.
If it was less than zero then the integral would not diverge to
positive infinity, there would be a negative sign appearing.\\

This helps us because it means that
$\int_{t}^{t_{0}}\frac{\dot{a}^{2}}{\ddot{a}}du$ is positive and as
such $\int_{t}^{t_{1}} a(u)du > 0$ which means that we have three
out of four conditions satisfied for a PIU/FIU. We need to check
$\bar{L}$. Set
\begin{eqnarray}
\bar{L}(\bar{T}) &=&
\frac{\bar{\Omega}''\bar{\Omega}}{\bar{\Omega}'^{2}}\\
&=& \frac{a(t)\left(\ddot{a}(t)a^{2}(t) + \dot{a}^{2}(t)a(t)
\right)}{\dot{a}^{2}(t)a^{2}(t)}\\
&=& \frac{\ddot{a}a}{\dot{a}^{2}} + 1
\end{eqnarray}
But because $\frac{\ddot{a}a}{\dot{a}^{2}} = \bar{\lambda}-1$ as
$t\to 0$ we see $\bar{L}\to \bar{\lambda}$ as
$t\to 0$.\\

This proves the theorem. $\Box$\\

Previously we only considered a PIU/FIU with a finite value of
$\bar{\lambda}$ but now we will examine what happens for
$\bar{\lambda} = +\infty$.
\begin{theorem}
An FRW model admits a PIU/FIU with $\bar{\lambda} = +\infty$ at
which the fluid flow is regular if and only if $\mathop
{\lim}\limits_{\bar{T} \to 0}\frac{\ddot{a}a}{\dot{a}^{2}}=+\infty$
and $\mathop {\lim}\limits_{t \to 0}a(t)\to +\infty$.
\end{theorem}
\emph{Proof}\\

We previously showed, similar to Ericksson \cite{Ericksson}, that if
$\bar{\lambda} = +\infty$ then $\mu\in\mathbb{R}^{+}\cup\{+\infty\}$
and
\begin{eqnarray}
p &=& -\frac{2}{3}\bar{L}\mu\left[1+ o(1)\right]\textrm{.}
\end{eqnarray}
Putting this into the following
\begin{eqnarray}
\dot{\mu} &=& -3(\mu + p)\frac{\dot{a}}{a}\\
&=&2\bar{L}\mu\left[1+ o(1)\right]\frac{\dot{a}}{a}
\end{eqnarray}
So it means that
\begin{eqnarray}
& &\frac{\dot{\mu}}{\mu} = 2\bar{L}\left[1+ o(1)\right]\frac{\dot{a}}{a}\\
& &\therefore 2\frac{\dot{a}}{a} =
o\left(\frac{\dot{\mu}}{\mu}\right)\\
& &\Longrightarrow 2\int_{t}^{t_{1}}\frac{\dot{a}(u)}{a(u)}du =
o\left(\int_{t}^{t_{1}}\frac{\dot{\mu}(u)}{\mu(u)}du \right)\\
& & \Longrightarrow 2\left(\ln a(t_{1}) - \ln a(t)\right) =
o\left(\ln\mu(t_{1}) -\ln\mu(t)\right)\\
& & -2\ln a(t) = o\left(\ln\mu(t)\right)\\
& & a^{-2} = o\left(\mu\right)\\
& & a^{-2}\mu^{-1} = o\left(1\right)
\end{eqnarray}
and hence $a^{2}\mu\to+\infty$, which again means that
$\dot{a}^{2}\to+\infty$. Just to be clear, we know that this is true
even if $\mu\in\mathbb{R}^{+}$ because $a\to+\infty$. We now
demonstrate the following
\begin{eqnarray}
\frac{\ddot{a}}{a} &=& -\frac{1}{6}\left(\mu + 3p\right)\\
&=& -\frac{1}{6}\left(\mu - 2\bar{L}\mu\left[1+ o(1)\right]\right)\\
&=& \frac{1}{3}\bar{L}\mu\left[1+ o(1)\right]\\
\Longrightarrow\frac{\ddot{a}a}{\dot{a}^{2}} &=&
\frac{1}{3}\bar{L}\mu\frac{a^{2}}{\dot{a}^{2}}\left[1+ o(1)\right]\\
&=&\frac{1}{3}\bar{L}\frac{3\dot{a}^{2} + 3K}{\dot{a}^{2}}\left[1+ o(1)\right]\\
&=&\bar{L}\left(1 + \frac{K}{\dot{a}^{2}}\right)\left[1+ o(1)\right]
\end{eqnarray}
and because $\dot{a}^{2}\to+\infty$ it is obvious that
$\frac{\ddot{a}a}{\dot{a}^{2}}$ diverges to positive infinity as
$\bar{T}\to 0$ because we have assumed $\bar{\lambda}\to+\infty$.\\

We now want to prove the converse. Consider an FRW model for which
$\frac{\ddot{a}a}{\dot{a}^{2}}\to+\infty$ as $t\to0$. We know that
$a(t)>0$ and $\dot{a}^{2}>0$ on a neighbourhood $(0,t_{1}]$ so this
means that there must exist a $t_{2} \in (0,t_{1}]$ such that
$\ddot{a}(t) > 0$ on $(0,t_{2}]$. So what this means is that
$\dot{a}(t)$ must be a strictly monotonically increasing function on
$(0,t_{2}]$ and hence as $t\to 0$, $\dot{a}\to \alpha \in
\mathbb{R}^{+}$  or
$\dot{a}\to +\infty$.\\

Let us take
\begin{eqnarray}
\frac{\ddot{a}a}{\dot{a}^{2}} &=& F(t)\qquad\textrm{where }
\Longrightarrow F(t)\to \infty\qquad\textrm{as } t\to 0\\
\Longrightarrow\frac{\dot{a}^{2}}{\ddot{a}a} &=& o(1)
\end{eqnarray}
So we know that
\begin{eqnarray}
\frac{1}{a} &=&o\left(\frac{\ddot{a}}{\dot{a}^{2}}
\right)\textrm{.}
\end{eqnarray}
Now consider
\begin{eqnarray}
\int_{t}^{t_{2}}\frac{\ddot{a}(u)}{\dot{a}^{2}(u)}du &=& \frac{1}{\dot{a}(t_{2})} - \frac{1}{\dot{a}(t)}
\end{eqnarray}
as we know that $\dot{a}\to\alpha\in\mathbb{R}^{+}\cup\{+\infty\}$
it is true that
\begin{eqnarray}
\int_{t}^{t_{2}}\frac{\ddot{a}(u)}{\dot{a}^{2}(u)}du &\to& \gamma \in \mathbb{R}^{+}
\end{eqnarray}
and hence that
\begin{eqnarray}
\int_{t}^{t_{1}}\frac{1}{a(u)}du &\to& \chi \in \mathbb{R}^{+}
\end{eqnarray}
So we know that, because of Lemma \ref{ErickssonLemma} and Corollary
\ref{ErickssonCorollary}, the FRW model will satisfy three
conditions for a PIU/FIU and we will prove the last condition now.
Set
\begin{eqnarray}
\bar{L}(\bar{T}) &=&
\frac{\bar{\Omega}''\bar{\Omega}}{\bar{\Omega}'^{2}}\\
&=& \frac{a(t)\left(\ddot{a}(t)a^{2}(t) + \dot{a}^{2}(t)a(t)
\right)}{\dot{a}^{2}(t)a^{2}(t)}\\
&=& \frac{\ddot{a}a}{\dot{a}^{2}} + 1
\end{eqnarray}
But because $\frac{\ddot{a}a}{\dot{a}^{2}} = +\infty$ as $t\to 0$ we
can say $\bar{L}\to \bar{\lambda} =
+\infty$ as $t\to 0$.\\

We have been able to prove the theorem. $\Box$\\

So what this theorem does, in conjunction with The FRW Result
\cite{Ericksson}, is prove that the FRW solution studied in this
paper (with $\eta_{0}\geq 1$) cannot admit an IPS/IFS or a PIU/FIU.
It cannot admit an IPS/IFS because $\mathop {\lim}\limits_{T \to
0}\frac{\ddot{a}a}{\dot{a}^{2}} \geq 0$ \cite{Ericksson} and it
cannot admit a PIU/FIU because $a(t)$ does not diverge as $t\to
0^{\pm}$.
\section{Conclusions and Further Outlook}
In this paper we have shown that there exist large classes of FRW
models that obey conformal definitions from the Quiescent Cosmology
framework. We were able to show that FRW models, whose scale factor
is represented as a generalised and finite series expansion, will
represent an FIU if that scale factor diverges (for $\eta_{0}<0$) at
a particular instant in time. When the scale factor vanishes, we
showed that some models represent an IPS/IFS (for
$\eta_{0}\in(0,1)$) and were able to prove that the remaining cases
(for $\eta_{0}\geq 1$) cannot represent an IPS/IFS or a PIU/FIU.\\

The FRW Result of Ericksson \cite{Ericksson} provides a simple
method of determining when an FRW solution can admit an IPS. In this
paper, we formed an equivalent statement for the FIU. In proving the
FRW Result, Ericksson also determined a gamma-law equation of state
for FRW models that represent an IPS. In the course of providing
relevant background information for The FRW (PIU/FIU) Result, we
were also able to demonstrate that there exists a gamma-law equation
of state for FRW solutions that admit an FIU. With the two FRW
results (for the IPS and FIU respectively), we now know it will
always be clear when an FRW model admits one of the isotropic
structures from Quiescent Cosmology.\\

The synthesis between the FRW and Quiescent Cosmology frameworks is
a natural one due to their respective ability at describing the
observable Universe. Quiescent Cosmology is attempting to describe
the very early Universe and hence the Big Bang while FRW models
offer reasonable explanations for both the Big Bang and the current
large scale state of the Universe. By studying both approaches to
cosmology, it may be possible to reveal even more information about
the Big Bang and the current Universe.\\

It is important to further our knowledge of FRW models that cannot
obey either of the isotropic definitions from Quiescent Cosmology as
they could provide insight for possible future alternate isotropic
structures. It is also important to understand the link between the
IPS/IFS and FIU, and the Big Bang/Crunch and Big Rip respectively -
this can be achieved by even more study of the overlap between the
FRW and Quiescent Cosmology frameworks.
\section*{References}
\bibliographystyle{unsrt}
\bibliography{Bibliography}

\begin{thebibliography}{10}

\bibitem{Lazkoz}
{L}. {F}ern\'andez {J}ambrina and {R}. {L}azkoz.
\newblock {G}eodesic {B}ehaviour around {C}osmological {M}ilestones.
\newblock {\em {J}ournal of {P}hysics}, 66:012015, 2007.

\bibitem{Visser}
{C}. {C}atto\"en and {M}. {V}isser.
\newblock {N}ecessary and {S}ufficient {C}onditions for {B}ig {B}angs,
  {B}ounces, {C}runches, {R}ips, {S}udden {S}ingularities and {E}xtremality
  {E}vents.
\newblock {\em {C}lassical and {Q}uantum {G}ravity}, 22:4913--4930, 2005.

\bibitem{Barrow}
{J}.{D}. {B}arrow.
\newblock {Q}uiescent {C}osmology.
\newblock {\em {N}ature}, 272:211--215, 1978.

\bibitem{Scott}
{P}.~{A}. {H}\"ohn and {S}.~{M}. {S}cott.
\newblock {E}ncoding {C}osmological {F}utures with {C}onformal {S}tructures.
\newblock {\em {C}lassical and {Q}uantum {G}ravity}, 26:035019, 2009.

\bibitem{Goode}
{S}.~{W}. {G}oode and {J}. {W}ainwright.
\newblock {I}sotropic {S}ingularities in {C}osmological {M}odels.
\newblock {\em {C}lassical and {Q}uantum {G}ravity}, 2:99--115, 1985.

\bibitem{Ellis}
{G}. {F}.~{R}. {E}llis.
\newblock {R}elativistic {C}osmology.
\newblock In {\em General {R}elativity and {C}osmology; {P}roceedings of the
  {I}nternational {S}chool of {P}hysics - {E}nrico {F}ermi, {V}arenna, {I}taly;
  {U}nited {S}tates; 30 {J}une-12 {J}uly 1969}, pages 104--182, 1971.

\bibitem{Caldwell1}
{R}.~{R}. {C}aldwell.
\newblock {A} {P}hantom {M}enace.
\newblock {\em {P}hysics {L}etters {B}}, 545:23--29, 2002.

\bibitem{Caldwell2}
{R}.~{R}. {C}aldwell, {M}. {K}amionkowski, and {N}.~{N}. {W}einberg.
\newblock {P}hantom {E}nergy: {D}ark {E}nery with w < -1 {C}auses a {C}osmic
  {D}oomsday.
\newblock {\em {P}hysical {R}eview {L}etters}, 91(071301):071301, 2003.

\bibitem{Barrow2}
{J}.{D}. {B}arrow.
\newblock {M}ore {G}eneral {S}udden {S}ingularities.
\newblock {\em {C}lassical and {Q}uantum {G}ravity}, 21:5619--5622, 2004.

\bibitem{Misner}
{C}.~{W}. {M}isner.
\newblock {T}ransport {P}rocesses in the {P}rimordial {F}ireball.
\newblock {\em {N}ature}, 214:40--41, 1967.

\bibitem{Scott2}
{S}.{M}. {S}cott.
\newblock {T}he {G}eneral {V}orticity {R}esult.
\newblock (In preparation).

\bibitem{Ericksson}
{G}. {E}ricksson.
\newblock {\em {T}he {I}sotropic {S}ingularity in {C}osmological {M}odels}.
\newblock PhD thesis, The Australian National University, 2003.

\end{thebibliography}
\end{document}